\documentclass[aps,12pt,nofootinbib]{revtex4}

\usepackage{epsfig}
\usepackage{graphicx,graphics}
\usepackage{amsmath}
\usepackage{amsfonts}
\usepackage{amssymb}

\begin{document}

\title{New Inequalities in Equilibrium Statistical Mechanics}

\author{J. G. Brankov$^{1,2}$ and N. S. Tonchev$^3$}

\affiliation{$^1$ Bogoliubov Laboratory of Theoretical Physics, Joint Institute for Nuclear Research, 141980 Dubna, Russia\\
$^2$ Institute of Mechanics, Bulgarian Academy of Sciences, 1113 Sofia, Bulgaria\\
$^3$ Institute of Solid State Physics, Bulgarian Academy of Sciences,
1784 Sofia, Bulgaria}



\maketitle



{\bf Abstract:}

Recently, new thermodynamic inequalities have been obtained, which set bounds on the quadratic fluctuations of intensive observables of statistical mechanical
systems in terms of the Bogoliubov - Duhamel inner product and some thermal average values. It was shown that several well-known inequalities in equilibrium statistical mechanics emerge as special cases of these results. On the basis of the spectral representation, lower and upper bounds on the one-sided fidelity susceptibility were derived in analogous terms. Here, these results are reviewed and presented in a unified manner. In addition, the spectral representation of the symmetric two-sided fidelity susceptibility is derived, and it is shown to coincide with the one-sided case. Therefore, both definitions imply the same
lower and upper bounds on the fidelity susceptibility.

\section{Introdiction}

A number of important results on the role of critical fluctuations in systems with broken symmetries have been obtained in the past by using  famous inequalities due to Bogoliubov, Mermin-Wagner, Griffits, among others, see, e.g. \cite{G72,DLS,GN01,R}. A remarkable property of this approach is that the results obtained are exact and cannot be inferred from any perturbation theory. The benefits of having exact statements about  systems with strongly
interacting particles are difficult to overestimate.

The article is structured as follows. We begin by introducing notations and basic definitions of the models under consideration. In Section II, we review the recently obtained inequalities, in which the central role is played by the Bogoliubov - Duhamel inner product. In Section III we give the definitions
of fidelity and its second derivatives: the one-sided and symmetric two-sided fidelity susceptibilities. The new results concerning the derivation
of a spectral representation for the symmetric two-sided fidelity susceptibilities are presented in Section IV. Since it turns out that the two spectral
representations are the same, we note that both definitions imply the same lower and upper bounds, see Section V. The paper closes with
some concluding remarks in Section VI.

We shall consider families of Hamiltonians depending on auxiliary external fields $\nu$ and $\nu^{\star}$ conjugate to the intensive
operators $A^{\dagger}$ and $A$, respectively,
\begin{equation}
{\mathcal H}(\nu)\equiv {\mathcal H}-|\Lambda|  (\nu A^{\dagger} + \nu^{\star} A).
\label{nu}
\end{equation}
The  quantum statistical mechanical models are defined initially in a finite region $\Lambda$ of
the $d$-dimensional Euclidean space $R^d$ or the $d$-dimensional integer lattice $Z^d$. By $|\Lambda|$ we denote the volume of $\Lambda$
in the former case, or the number of lattice sites in the latter case.

The Hamiltonian ${\mathcal H}(\nu)$ is defined as a self-adjoint operator in a separable Hilbert space {\bf H}, and the
corresponding free energy density $f_{\Lambda}[{\mathcal H}(\nu)]$ is assumed to exist. For the sake of simplicity, we
do not explicitly distinguish between a Hamiltonian ${\mathcal H}(\nu)$, describing a system with fixed number of particles
$N$ in $\Lambda$, and the statistical operator ${\mathcal H}(\nu) - \mu {\mathcal N}$ in the grand canonical ensemble,
where $\mu$ is the chemical potential and ${\mathcal N}$ is the particle number operator. The density of the corresponding thermodynamic
potential is given by
\begin{equation}
f_{\Lambda}[\mathcal{H}(\nu)] = - (\beta |\Lambda|)^{-1}\ln
Z_{\Lambda}[\mathcal{H}(\nu)] \label{1c3},
\end{equation}
where $Z[\mathcal{H}(\nu)]:=\mathrm{Tr}{\mathrm e}^{-\beta \mathcal{H}(\nu)}$ is the partition function.
Average values in the Gibbs ensemble with the Hamiltonian ${\mathcal H}(\nu)$ are defined as
\begin{equation}
\langle \cdots \rangle_{\mathcal{H}(\nu)} \equiv
\mathrm{Tr}({\mathrm e}^{-\beta \mathcal{H}(\nu)}\cdots\,)
/Z_{\Lambda}[\mathcal{H}(\nu)].
\end{equation}

In the theory of phase transitions a key role play the following quantities:
the quadratic fluctuations \begin{equation}
\langle \delta A \delta A^{\dagger}
\rangle_{{\mathcal H}(\nu)}, \quad \delta A \equiv A -\langle A\rangle_{{\mathcal H}(\nu)},
\label{kf}
\end{equation}
and the second derivative of the free energy density
with respect to external fields $\nu$, $\nu^{\star}$, i.e., the isothermal susceptibility
\begin{equation}
\chi_{\Lambda}(\nu) = -\frac{\partial^2 f_{\Lambda}[{\mathcal
H}(\nu)]}{\partial \nu ^\star \partial\nu}.
\label{fed}
\end{equation}
In what follows we will focus  on the problem of relating (\ref{kf}) and (\ref{fed})
to other quantities that appear as indicators of phase transitions. Much work have been done on this issue in the context of quantum information theory (see, e.g.,  \cite{Gu10, BT12} and references therein). As a result, the knowledge obtained so far shows that techniques borrowed from the quantum information theory
are likely to be useful in studying the phenomenon of phase transitions.

First of all we make use of the important relationship between (\ref{fed}) and
the Bogoliubov - Duhamel inner product, defined as
\begin{equation}
(A; B)_{\mathcal H} \equiv (Z [{\mathcal H}])^{-1}\int_0^{1}
{\mathrm d}\tau \: \mathrm{Tr} \left[{\mathrm e}^{-\beta(1
-\tau){\mathcal H}} A^{\dagger} {\mathrm e}^{-\beta \tau {\mathcal
H}} B\right] \label{1M},
\end{equation}
namely,
\begin{equation}
-\frac{1}{\beta |\Lambda|}\frac{\partial^2 f_{\Lambda}[{\mathcal
H}(\nu)]}{\partial \nu ^\star \partial\nu}=(\delta A;\delta A).
\label{second}
\end{equation}
This object  has been introduced and discussed by a number of authors (see \cite{DLS, BT11} and references therein), which noted that it is
useful in finding quantum generalizations of different
inequalities in statistical mechanics of classical systems. Let us note that
if the operator $A$ commutes with the Hamiltonian ${\mathcal H}$ (commutative case)
the expressions (\ref{kf}) and (\ref{second}) coincide. The genuine difference  between (\ref{kf}) and (\ref{second}) in the noncommutative case
constitutes the novelty of the inequalities obtained below.

In the remainder we consider given observables $A, B, C, \dots$ and fixed Hamiltonian ${\mathcal H}$ pertaining to a quantum
system in a finite region $\Lambda$. Whenever no confusion arises, for brevity of notation we will omit the subscripts
${\mathcal H}$ and $\Lambda$, as well as the argument of the partition function $Z$.

\section{Bounds on the quadratic fluctuations}

We assume that the Hamiltonian ${\mathcal H}$ of the system in
a finite region of space $\Lambda$ is
a self-adjoint, trace-class operator which generates the Gibbs semigroup
$\{\exp(-\beta {\mathcal H}_{\Lambda})\}_{\beta \ge 0}$. Let the operators
$A$, $B,\dots$, belong to the algebra of bounded observables, for which
the Bogoliubov - Duhamel inner product $(A;B)$ is well defined.

Further we assume that the Hamiltonian ${\mathcal H}$ has a
discrete, non-degenerate spectrum only, $\{E_n, n=1,2,3\dots \}$ and denote
by $|n\rangle$ the corresponding eigenfunctions, i.e., ${\mathcal
H}|n\rangle = E_n |n\rangle$, $n=1,2,3\dots$. By $A_{mn}=\langle
m|A|n\rangle$ we denote the corresponding matrix element of an
operator $A$. Then, the spectral representation of the Bogoliubov - Duhamel inner
(\ref{1M}) can be written as
\begin{equation}
(A ;B)_{{\mathcal H}} = (Z_{\Lambda}
[{\mathcal H}])^{-1}\sum_{m,n}{}^\prime A^{*}_{mn}B_{mn}
\frac{e^{-\beta E_{m}} - e^{-\beta E_{n}}}
{\beta(E_{n}-E_{m})} + (Z_{\Lambda}
[{\mathcal H}])^{-1}\sum_{n}{\mathrm e}^{-\beta E_{n}} A^{*}_{nn}B_{nn},
\label{M5}
\end{equation}
where the prime in the double sum means that the term with $n=m$ is
excluded.

Our aim is to majorize the quadratic fluctuations
\begin{equation}
\langle \delta A^{\dagger}\delta A\rangle =
\langle A^{\dagger} A\rangle_{\mathcal H} -|\langle A\rangle_{\mathcal H}|^2
\label{Fluc3}
\end{equation}
by terms proportional to some power of the inner product
\begin{eqnarray}
&&(\delta A; \delta A)= (A;A)- |\langle A\rangle|^2 \nonumber \\ && =
Z_{\Lambda}^{-1}
\sum_{m,n}{}^\prime |A_{mn}|^2\frac{{\mathrm e}^{-\beta E_{m}} -
{\mathrm e}^{-\beta E_{n}}}
{\beta(E_{n}-E_{m})}+ (Z_{\Lambda}[{\mathcal H}])^{-1}
\sum_{n}{\mathrm e}^{-\beta E_n} |A_{nn}|^2 -|\langle A\rangle|^2.
\label{Fluc2}
\end{eqnarray}

Fot the symmetrized form of (\ref{Fluc3}) we obtain
\begin{eqnarray}
&&\frac{1}{2}\langle A^{\dagger} A + A A^{\dagger}\rangle - (A; A)
\nonumber \\&& = Z^{-1}
\sum_{m,n}{}^\prime |A_{mn}|^2\left\{\frac{1}{2}({\mathrm e}^{-\beta E_{n}}
+ {\mathrm e}^{-\beta E_{m}})
- \frac{{\mathrm e}^{-\beta E_{m}} - {\mathrm e}^{-\beta E_{n}}}
{\beta(E_{n}-E_{m})}\right\},
\label{dif}
\end{eqnarray}
which, by using the identity
\begin{equation}
{\mathrm e}^{-\beta E_{m}} + {\mathrm e}^{-\beta E_{n}}=
({\mathrm e}^{-\beta E_{n}}-{\mathrm e}^{-\beta E_{m}})\coth
\frac{\beta(E_{m}-E_{n})}{2}
\label{coth1}
\end{equation}
can be expressed as:
\begin{equation}
\frac{1}{2}\langle A^{\dagger} A + A A^{\dagger}\rangle - (A; A)
 = Z^{-1}
\sum_{m,n}{}^\prime |A_{mn}|^2\frac{{\mathrm e}^{-\beta E_{m}} -
{\mathrm e}^{-\beta E_{n}}}
{\beta(E_{n}-E_{m})}(X_{mn}\coth X_{mn} -1),
\label{dif1}
\end{equation}
where $X_{mn}= \beta (E_{m}-E_{n})/2$.

Different choices of the  upper bound on the right-hand side of
(\ref{dif1}) generate different inequalities. Thus, the
 inequality of Brooks Harris \cite{Harris},
\begin{equation}
(A;A)\leq \frac{1}{2}\langle AA^+ + A^+A\rangle \leq
(A;A)+\frac{\beta}{12}\langle [[A^+,{\mathcal H}],A]\rangle .
\label{Harris}
\end{equation}
is obtained by setting $1\leq x\coth x \leq 1+x^2/3$.

On the other hand, if one uses another elementary inequality,
$1\leq x\coth x \leq 1+|x|$,
and subsequently  applies the H\"{o}lder inequality, one obtains
the result due to Ginibre \cite{G}:
\begin{equation}
(A;A)\leq \frac{1}{2}\langle AA^{+} + A^{+}A\rangle\leq (A;A)+
\frac{1}{2}\{(A;A)\beta \langle [[A^+,{\mathcal
H}],A]\rangle\}^{\frac{1}{2}} \label{G}.
\end{equation}

A different choice of the parameters in the H\"{o}lder inequality,
followed by the implementation of the upper bound
\begin{equation}
|\ e^{-\beta E_{l}} - e^{-\beta E_{m}}| < |\ e^{-\beta E_{l}} +
e^{-\beta E_{m}}|, \label{ub}
\end{equation}
generates a symmetric version of the inequality due to Bogoliubov
Jr. \cite{Bog72}:
\begin{equation}
\frac{1}{2}\langle AA^+ + A^+A\rangle \leq
(A;A)+\frac{1}{2}[(A;A)\beta]^{2/3}\{\langle [A^+,{\mathcal
H}][{\mathcal H}, A]+ [{\mathcal H}, A][A^+,{\mathcal
H}]\rangle\}^{1/3}, \label{Jr}
\end{equation}

To derive generalizations of the known inequalities involving the
Bogoliubov - Duhamel  inner product, we define a set of new functionals
in terms of their spectral representations:
\begin{eqnarray}
&&F_{2n}(J;J) \equiv Z^{-1} \sum_{ml}|J_{ml}|^{2}|e^{-\beta E_{l}}
- e^{-\beta E_{m}}|(\beta|E_{m}-E_{l}|)^{2n-1}
\nonumber\\&&=\beta^{2n}(R_{n};R_{n})=
\beta^{2n-1}\langle[R_{n}^{+}R_{n-1}-R_{n-1}R_{n}^{+}]\rangle, \quad n=0,1,2,3,\dots,
\label{BT10}
\end{eqnarray}
where, by definition, $R_{-1}\equiv X_{J{\mathcal H}}$ is a
solution of the operator equation $J = [X_{J{\mathcal
H}},{\mathcal H}]$, and
\begin{equation}
R_{0}\equiv R_{0}(J)= J,\quad  R_{1}\equiv R_{1}(J) =[{\mathcal
H},J],\; \dots, \; R_{n} \equiv R_{n}(J)= [{\mathcal H},R_{n-1}(J)].
\label{R}
\end{equation}
The observables $R_k$, $k=0,1,2,\dots$, were introduced in \cite{BPR}. Next,
we have defined
\begin{eqnarray}
&& F_{2n+1}(J;J)\equiv  Z^{-1}\sum_{ml}|J_{ml}|^{2}(e^{-\beta
E_{l}} + e^{-\beta E_{m}})[\beta(E_{m}-E_{l})]^{2n}
\nonumber\\&&=\beta^{2n}\langle[R_{n}R_{n}^{+}+R_{n}^{+}R_{n}]\rangle, \quad n=0,1,2,3,\dots.
\label{M911}
\end{eqnarray}
In particular,
\begin{eqnarray}
&&F_{0}(J;J)= (J;J), \quad F_{1}(J;J)= \langle JJ^+ + J^+J\rangle,
\quad F_{2}(J;J)= \beta \langle [[J^+,{\mathcal H}],J]\rangle,
\nonumber \\&& F_{3}(J;J)= \beta^2 \langle [J^+,{\mathcal
H}][{\mathcal H}, J]+ [{\mathcal H}, J][J^+,{\mathcal H}]\rangle.
\label{Part}
\end{eqnarray}
The functionals (\ref{BT10}) and (\ref{M911}) are used to generalize all the
known inequalities used in the Approximating Hamiltonian method.
Here we give the final results:

\subsection{The generalized Harris inequality}

For all integers $n=0,1,2,\dots$ we have
\begin{equation}
F_{2n}(J;J)\leq \frac{1}{2}F_{2n+1}(J;J)\leq
F_{2n}(J;J)+\frac{1}{12}F_{2n+2}(J;J) . \label{BHgen}
\end{equation}

The Brooks Harris inequality
(\ref{Harris}) is recovered when $n=0$.

\subsection{The generalized Plechko inequalities}

The following inequalities were proved to hold for all $p,q > 1$,
such that $1/p + 1/q = 1$:
\begin{eqnarray}
&&(2Z)^{-1}\sum_{ml}|J_{ml}|^{2}|e^{-\beta E_{l}} - e^{-\beta
E_{m}}|[\beta(E_{m}-E_{l})]^{2n}\nonumber \\ &&\leq
\frac{1}{2}(J;J)^{1/p}\left\{Z^{-1}\sum_{ml}|J_{ml}|^2 \frac
{e^{-\beta E_{l}} - e^{-\beta E_{m}}}
{\beta(E_{m}-E_{l})}[\beta|E_{m}-E_{l}|]^{(2n+1)q} \right\}^{1/q}.
\label{estH1}
\end{eqnarray}

One of the possible choices of $p$ and $q$ here is even
integer $q=2k$ (hence, $p= 2k/(2k-1)$) which leads to the set of
generalized Ginibre inequalities.

\subsection{The generalized Ginibre inequalities}

These inequalities read ($k=1,2,3,\dots$):
\begin{equation}
F_{2n}(J;J)\leq \frac{1}{2}F_{2n+1}(J;J)\leq F_{2n}(J;J)+
\frac{1}{2}(J;J)^{(2k-1)/2k}[F_{2k(2n+1)}(J;J)]^{1/2k}.
\label{Ggen}
\end{equation}

At $n=0$ the above set reduces to a symmetric version of the
inequalities obtained by Plechko \cite{P}:
\begin{equation}
(J;J)\leq \frac{1}{2}\langle JJ^{+} +J^{+}J\rangle \leq (J;J)+
\frac{1}{2}(J;J)^{(2k-1)/2k}\beta (R_k;R_k)^{1/2k}, \quad
(k=1,2,3,\dots) . \label{Plechko}
\end{equation}
Hence, in the particular case of $k=1$ one obtains the Ginibre
inequality (\ref{G}).

\subsection{The generalized Bogoliubov Jr. - Plechko - Repnikov inequalities}

These inequalities are obtained from (\ref{estH1}) under the choice of odd $q=2k+1$, hence,
$p=(2k+1)/2k$, $k=1,2,3,\dots$:
\begin{equation}
\frac{1}{2}F_{2n+1}(J;J)\leq
F_{2n}(J;J)+\frac{1}{2}(J;J)^{2k/(2k+1)}[F_{2(2nk+n+k)+1}(J;J)]^{1/(2k+1)}.
\label{genJr}
\end{equation}

At $n=0$ these reduce to a symmetric version of the set of
inequalities obtained by Bogoliubov Jr., Plechko and Repnikov
\cite{BPR}:
\begin{equation}
\frac{1}{2}\langle JJ^+ + J^+J\rangle \leq
(J;J)+\frac{1}{2}(J;J)^{2k/(2k+1)}\{\beta^{2k}\langle R_k R_k^{+}
+ R_k^{+} R_k\rangle\}^{1/(2k+1)}. \label{BPR}
\end{equation}

The symmetric version of the inequality due to Bogoliubov Jr.
(\ref{Jr}) follows from here in the particular case of $k=1$.

In \cite{BT11} we have shown that under
sufficiently mild conditions, each of the generalized upper bounds has the same form and order of magnitude with
respect to the number of particles (or volume) for all the quantities derived by
commutations of an intensive observable with the Hamiltonian of the system.
An application of the generalized inequalities to a quantum spin model
with separable attraction and the Dicke model of superradiance was
given too.

\section{Fidelity susceptibility and Gibbs thermal states}

Over the last decade there have been impressive theoretical advances concerning
the concepts of entanglement and fidelity from quantum and information theory
\cite{BZ06},\cite{AFOV08}, and  their application in condensed matter physics, especially in
the theory of critical phenomena and phase transitions, for a review see \cite{AF09,Gu10}.
These two concepts are closely related to each other.

The  fidelity \cite{J94,Sh95} naturally appears in quantum mechanics as the
absolute value of the overlap (Hilbert-space scalar product) of two quantum states corresponding to different
values of the control parameters.
 The corresponding finite-temperature extension, defined as a
functional of two density matrices, $\rho_1$ and $\rho_2$,
\begin{equation}
{\cal F}(\rho_1,\rho_2) = \mathrm{Tr}\sqrt{\rho_1^{1/2} \rho_2 \rho_1^{1/2}}, \label{defFidel}
\end{equation}
was introduced by Uhlmann \cite{U76} and called fidelity by Jozsa \cite {J94}.

Being a measure of the similarity between quantum states, both pure or mixed, fidelity should decrease abruptly at a critical point, thus locating and characterizing the phase transition. Different finite-size scaling behaviors of the fidelity indicate different types of phase transitions.
The fidelity approach is basically a metric one and has an advantage over the traditional
Landau-Ginzburg theory, because it avoids possible difficulties in identifying the notions of order parameter, symmetry breaking, correlation length and so it is suitable for the study of different kinds of topological  or
Berezinskii-Kosterlitz-Thouless  phase transitions.

The above mentioned decrease in the fidelity ${\cal F}(\rho_{1}, \rho_{2})$, when the state $\rho_{2}$ approaches a quantum critical state $\rho_{1}$, is associated with a divergence of the fidelity susceptibility $\chi_F(\rho_{1})$ which reflects the singularity of ${\cal F}(\rho_{1}, \rho_{2})$ at that point.
The fidelity susceptibility $\chi_F(\rho_{1})$, which is the main objects of this study,
naturally arises as a leading-order term
in the expansion of the fidelity for two infinitesimally close density matrices $\rho_{1}$ and $\rho_{2}=
\rho_{1} + \delta \rho$. For simplicity, in this section  and thorough the rest of the paper we set in (\ref{nu})
$|\Lambda|  (\nu A^{\dagger} + \nu^{\star} A)=hS$, where $h=\nu=\nu ^{\star}$ is a real parameter. Following our study \cite{BT12}, we consider the one-parameter family of Gibbs states
\begin{equation}
\rho(h) = [Z(h)]^{-1}\exp[-\beta T +\beta h S], \label{roh}
\end{equation}
defined on the family of Hamiltonians of the form $H(h) = T - h S$,
where the Hermitian operators $T$ and $S$ do not commute in the
general case, $h$ is a real parameter, and $Z(h)= {\mathrm Tr}\exp[-\beta T +\beta h S]$
is the corresponding partition function. There are two natural definitions of the fidelity susceptibility
at the point $h=0$, depending on the type of approach to that point:

(i) The one-sided fidelity susceptibility at the
point $h=0$ in the parameter space is defined as (see e.g. \cite{V10}):
\begin{equation}
\chi_{F}(\rho(0)):=\lim _{h \rightarrow 0}\frac{-2\ln {\cal F}(\rho(0),\rho(h))}{h^{2}}=
-\left. \frac{\partial^{2}{\cal F}(\rho(0),\rho(h))}{\partial h^{2}}\right|_{h=0} .
\label{dchi}
\end{equation}
Starting from this definition, in our work \cite{BT12} we derived a spectral representation for
$\chi_{F}(\rho(0))$ which was used to obtain lower and upper bounds on the fidelity susceptibility.

(ii) One may consider  also a symmetric two-sided definition of the fidelity susceptibility at the
point $h=x \in R$, given by
\begin{equation}
\chi_{F}^{(2)}(\rho(x)):=\lim _{h \rightarrow 0}\frac{-2\ln {\cal F}(\rho(x-h/2),\rho(x+h/2))}{h^{2}}=
-\left. \frac{\partial^{2}{\cal F}(\rho(x-h/2),\rho(x+h/2))}{\partial h^{2}}\right|_{h=0}.
\label{dchi2}
\end{equation}

To avoid confusion, we point out that for mixed states the definition of the fidelity susceptibility
(\ref{dchi}), based on the Uhlmann fidelity (\ref{defFidel}), differs from the one derived in \cite{AASC10}
(see also \cite{Gu10}) by extending the ground-state Green's function representation to nonzero temperatures, even both have the same $T=0$ limit.
This fact has been pointed out in \cite{S10}, see also our discussion in \cite{BT12}. Along with the statistical mechanical notion of susceptibility,
the quantity (\ref{dchi}) is known also as ``Bures metric over the thermal states'' \cite{AASC10,ZVG07}, thus introducing the geometric
approach in the field.

To proceed with the calculations when the operators
$T$ and $S$ do not commute, one has to consider a convenient spectral representation.
To simplify the problem, we assume that the Hermitian operator $T$ has a complete
orthonormal set of eigenvectors $|n\rangle$, $T|n\rangle = T_n|n\rangle$, where $n=1,2,\dots $,
with non-degenerate spectrum $\{T_n\}$.

In the one-sided case (\ref{dchi}), the following spectral representation was obtained
\cite{BT12}:
\begin{eqnarray}
\chi_F(\rho) &=&  \frac{1}{2} \sum_{m,n}
\frac{|\langle m|\rho'(0)|n\rangle|^2}{\rho_m + \rho_n} \nonumber \\
&=&\frac{\beta^2}{8}\sum_{m,n, m\not=n} \frac{\rho_n
-\rho_m}{X_{mn}} \frac{|\langle n|S|m \rangle|^2}{X_{mn}\coth
X_{mn}} + \frac{1}{4}\beta^2 \langle (\delta S^d)^2\rangle_0 .
\label{FiSus2}
\end{eqnarray}
Here $X_{mn} \equiv  \beta (T_m -T_n)/2$, $\langle \cdots \rangle_0$ denotes the Gibbs average value at
$h=0$, $\delta S^d = S^d - \langle S^d\rangle_0$, where $S^d$ is the diagonal part of the operator $S$,
so that
\begin{equation}
\langle (\delta S^d)^2 \rangle_0 :=\sum_m \rho_{m}\langle m|S|m \rangle^2 - \langle S\rangle_0^2.
\end{equation}

Representation (\ref{FiSus2}) was the starting point for the derivation of inequalities involving macroscopic quantities, like susceptibilities and thermal average values. Note that the first term in the right-hand side describes the purely quantum contribution to
the fidelity susceptibility, which vanishes when the operators $T$ and $S$ commute,
while the second term represents the  ``classical'' contribution.

By comparing definitions (\ref{dchi}) and (\ref{dchi2}) in the non-commutative case, one sees an essential
difference: in the one-sided case the zero-field density matrix $\rho(0)$ is diagonal in the basis
spanned by the eigenfunctions of the operator $T$, $\langle m|\rho(0)|n\rangle = \rho_m(0) \delta_{m,n}$, $m,n =1,2,\dots $, while in the two-sided case (\ref{dchi2}) both $\rho(-h/2)$ and $\rho(h/2)$ have non-diagonal
elements. Hence, it is not obvious that the two-sided fidelity susceptibility will have the same spectral
representation (\ref{FiSus2}) and, the following from it, lower and upper bounds. This observation has motivated
us to give an independent derivation of the spectral representation for $\chi_{F}^{(2)}(\rho(0))$.

\section{Derivation of the spectral representation for the symmetric two-sided fidelity susceptibility}

For the family of density matrices of the form (\ref{roh}) we consider the slightly more general case of a
two-sided fidelity around a point $x \in R$:
\begin{equation}
{\cal F}(\rho(x-y), \rho(x+y)) = \mathrm{Tr}\sqrt{\rho^{1/2}(x-y) \rho(x+y) \rho^{1/2}(x-y)}, \qquad y \in R.
\label{Fidel2}
\end{equation}
Assume that $\{|n\rangle,\; n=1,2,\dots\}$ is a complete set of eigenvectors of the Hamiltonian $H(x) =
T -xS$, with eigenvalues $E_n(x)$: $(T -xS)|n\rangle = E_n(x)|n\rangle$. In this basis the density matrix
$\rho(x)$ is diagonal too, which implies $\langle m|\rho(x)|n\rangle = \rho_n(x) \delta_{m,n}$, for all $m,n=1,2,\dots $.

In order to calculate the symmetric two-sided fidelity susceptibility at $y=0$, we need the
$O(h^2)$ term in the expansion of (\ref{Fidel2}) in powers of $h \rightarrow 0$. Consider first the
expansions
\begin{equation}
\rho(x\pm y)= \rho(x) \pm y \rho'(x) + \frac{1}{2}y^2 \rho''(x) + r_{\pm},
\label{exppm}
\end{equation}
where $r_{\pm}$ are bounded operators of the order $O(h^3)$. From the normalization condition for the density
matrices, $\mathrm{Tr}\rho(x\pm y)= \mathrm{Tr}\rho(x)=1$, it follows that
\begin{equation}
\mathrm{Tr}\rho'(x) = \mathrm{Tr} \rho''(x) =0.
\label{zerotr}
\end{equation}

To calculate the square root of $\rho(x-y)$ up to the order $O(h^2)$, we set
\begin{equation}
\rho^{1/2}(x\pm y)= \rho^{1/2}(x) \pm A + B +C_{\pm},
\label{sqrt}
\end{equation}
where $A$, $B$, and $C_{\pm}$ are bounded operators of the order of $y$, $y^2$, and $y^3$, respectively. Then, by comparing the squared of $(\ref{sqrt})$ with the expansion $(\ref{exppm})$ for $\rho(x-y)$, we
obtain
\begin{eqnarray}
 y \rho'(x)& = &\rho^{1/2}(x) A + A \rho^{1/2}(x) , \nonumber \\
\frac{1}{2}y^2 \rho''(x) &=& A^2 + \rho^{1/2}(x) B + B \rho^{1/2}(x) .
\end{eqnarray}
Hence, we find the matrix elements of the operators $A$ and $B$:
\begin{eqnarray}
\langle m|A|n \rangle &=& y \frac{\langle m|\rho'(x)|n \rangle}{\rho^{1/2}_m (x) + \rho^{1/2}_n (x)} ,
\label{Amn} \\
\langle m|B|n \rangle   &=& - \frac{\langle m|A^2|n \rangle}{\rho^{1/2}_m (x) + \rho^{1/2}_n (x)}  + \frac{1}{2}y^2 \frac{\langle m|\rho''(x)|n \rangle}{\rho^{1/2}_m (x) + \rho^{1/2}_n (x)}  .
\label{Bmn}
\end{eqnarray}

Next, with the aid of expansions (\ref{exppm}) and (\ref{sqrt}) we evaluate, up to order $O(y^2)$, the product
\begin{eqnarray}
&&\rho^{1/2}(x-y) \rho(x+y) \rho^{1/2}(x-y) = \rho^2(x) -A\rho^{3/2}(x)-\rho^{3/2}(x)A+ y\rho^{1/2}(x)\rho'(x)
\rho^{1/2}(x) \nonumber \\&& + A\rho(x)A
-y A \rho'(x)\rho^{1/2}(x)- y \rho^{1/2}(x)\rho'(x)A +\frac{1}{2}y^2\rho^{1/2}(x)\rho'(x)\rho^{1/2}(x)
\nonumber \\&& + B \rho^{3/2}(x)+ \rho^{3/2}(x)B + O(y^3).
\label{rrr}
\end{eqnarray}

Next, following the standard scheme proposed in \cite{SZ03}, we set
\begin{equation}
\sqrt{\rho^{1/2}(x-y) \rho(x+y) \rho^{1/2}(x-y)} = \rho(x) +X+Y + Z,
\label{sqrtrrr}
\end{equation}
where $X$, $Y$, and $Z$ are bounded operators of the order of $y$, $y^2$, and $y^3$, respectively,
and compare the terms of the same order of magnitude in the right-hand sides of (\ref{rrr}) and the
squared of (\ref{sqrtrrr}). Within the order $O(y)$ we obtain
\begin{equation}
\rho(x)X + X \rho(x) = -A \rho^{3/2}(x) - \rho^{3/2}(x)A + y \rho^{1/2}(x)\rho'(x)\rho^{1/2}(x)
\label{compy}
\end{equation}
By taking matrix elements with the eigenvectors of $H(x)$ and using expression (\ref{Amn}) for
$\langle m|A|n \rangle$, we find
\begin{eqnarray}
&&[\rho_m(x) + \rho_n(x)] \langle m|X|n \rangle = \nonumber \\&&-y\frac{\langle m|\rho'(x)|n \rangle}
 {\rho_m^{1/2}(x) + \rho_n^{1/2}(x)}\left[\rho_m^{3/2}(x) + \rho_n^{3/2}(x)\right] +
y \rho_m^{1/2}(x)\rho_n^{1/2}(x)\langle m|\rho'(x)|n \rangle =\nonumber \\
&& -y \langle m|\rho'(x)|n \rangle [\rho_m^{1/2}(x) - \rho_n^{1/2}(x)]^2.
\label{mXn}
\end{eqnarray}
Therefore,
\begin{equation}
\langle m|X|n \rangle = -y \frac{\langle m|\rho'(x)|n \rangle}{\rho_m(x) + \rho_n(x)}
[\rho_m^{1/2}(x) - \rho_n^{1/2}(x)]^2.
\label{mXn2}
\end{equation}
One important consequence of this equality is the vanishing of the diagonal elements of $X$, hence
$\mathrm{Tr}X =0$.

Next, within the order $O(y^2)$, from (\ref{rrr}) and the square of expression (\ref{sqrtrrr}) we obtain
\begin{eqnarray}
&&X^2 + \rho(x)Y + Y \rho(x) =
A\rho(x)A -y A \rho'(x)\rho^{1/2}(x)- y \rho^{1/2}(x)\rho'(x)A \nonumber \\ && +\frac{1}{2}y^2\rho^{1/2}(x)\rho''(x)\rho^{1/2}(x)
 + B \rho^{3/2}(x)+ \rho^{3/2}(x)B.
\label{compy2}
\end{eqnarray}
For our needs it suffices to take the diagonal elements of the above equality
\begin{eqnarray}
&&\langle n|X^2|n \rangle  + 2\rho_n(x)\langle n|Y|n \rangle =
\langle n|A\rho(x)A|n \rangle  -y \rho_n^{1/2}(x)\langle n|A \rho'(x) + \rho'(x)A|n \rangle  \nonumber \\ && +\frac{1}{2}y^2\rho_n(x)\langle n|\rho''(x)|n \rangle
 + 2\rho_n^{3/2}(x)\langle n|B|n \rangle ,
\label{nn2}
\end{eqnarray}
and evaluate $\mathrm{Tr}Y$. Since the calculations are rather involved, we present them in some detail. Taking into
account Eqs. (\ref{Amn}), (\ref{Bmn}) and  (\ref{mXn2}), we obtain the expression
\begin{eqnarray}
&&\mathrm{Tr}Y \equiv \sum_n \langle n|Y|n \rangle = -\frac{1}{2}\, y^2 \sum_{m,n} |\langle n|\rho'(x)|m \rangle|^2
\frac{[\rho_m^{1/2}(x)-\rho_n^{1/2}(x)]^4}{\rho_n(x)[\rho_m(x)+\rho_n(x)]^2}\nonumber \\ &&
+ \frac{1}{2}\, y^2 \sum_{m,n}
\frac{|\langle n|\rho'(x)|m \rangle|^2\rho_m(x)}{\rho_n(x)[\rho_m^{1/2}(x)+\rho_n^{1/2}(x)]^2} -
y^2\sum_{m,n}  \frac{|\langle n|\rho'(x)|m \rangle|^2}{\rho_n^{1/2}(x)[\rho_m^{1/2}(x)+\rho_n^{1/2}(x)]}\nonumber \\ &&
+\frac{1}{2}\, y^2 \mathrm{Tr}\rho''(x) - \frac{1}{2}\, y^2 \sum_{m,n}\frac{|\langle n|\rho'(x)|m \rangle|^2}
{[\rho_m^{1/2}(x)+\rho_n^{1/2}(x)]^2}.
\label{TrY1}
\end{eqnarray}
Now we note that the forth term in the right-hand side of the above equality is zero due to (\ref{zerotr}). The sum of the second and fifth
terms yields
\begin{equation}
\frac{1}{2}\, y^2\sum_{m,n}  \frac{|\langle n|\rho'(x)|m \rangle|^2[\rho_m(x)-\rho_n(x)]}{\rho_n(x)[\rho_m^{1/2}(x)+\rho_n^{1/2}(x)]^2}=
\frac{1}{2}\, y^2\sum_{m,n}  \frac{|\langle n|\rho'(x)|m \rangle|^2[\rho_m^{1/2}(x)-\rho_n^{1/2}(x)]}{\rho_n(x)[\rho_m^{1/2}(x)+\rho_n^{1/2}(x)]}.
\label{25}
\end{equation}
Next, by writing $[\rho_m^{1/2}-\rho_n^{1/2}]^4= (\rho_m + \rho_n)^2 - 4\rho_m^{1/2}\rho_n^{1/2} (\rho_m + \rho_n) + 4 \rho_m \rho_n$, we split
the first sum in the right-hand side of (\ref{TrY1}) into a sum of three terms:
\begin{equation}
-\frac{1}{2}\, y^2\sum_{m,n} |\langle n|\rho'(x)|m \rangle|^2 \left[\frac{1}{\rho_n(x)} - \frac{4\rho_m^{1/2}(x)}{\rho_n^{1/2}(x)
[\rho_m(x)+\rho_n(x)]} + \frac{4\rho_m(x)}{[\rho_m(x)-\rho_n(x)]^2}\right].
\label{1abc}
\end{equation}
By adding up (\ref{25}) with the third term in (\ref{TrY1}) and the first term coming from (\ref{1abc}) we obtain
\begin{equation}
-2 y^2\sum_{m,n} \frac{|\langle n|\rho'(x)|m \rangle|^2}{\rho_n^{1/2}(x)
[\rho_m^{1/2}(x)+\rho_n^{1/2}(x)]}.
\label{25a}
\end{equation}
Remarkably, the sum of the above result with the second term coming from (\ref{1abc}) vanishes:
\begin{equation}
2 y^2\sum_{m,n} \frac{|\langle n|\rho'(x)|m \rangle|^2[\rho_m^{1/2}(x)-\rho_n^{1/2}(x)]}
{[\rho_m^{1/2}(x)+\rho_n^{1/2}(x)][\rho_m(x)+\rho_n(x)]} = 0.
\label{25ab}
\end{equation}
Therefore, we are left with the contribution of the last term in (\ref{1abc}), hence
\begin{equation}
\mathrm{Tr}Y = - 2 y^2\sum_{m,n} \frac{|\langle n|\rho'(x)|m \rangle|^2\rho_m(x)}
{[\rho_m(x)+\rho_n(x)]^2} = - y^2\sum_{m,n} \frac{|\langle n|\rho'(x)|m \rangle|^2}
{[\rho_m(x)+\rho_n(x)]}.
\label{TrYfin}
\end{equation}

Thus, taking trace of both sides of equality (\ref{sqrtrrr}), we obtain the two-sided symmetric fidelity (\ref{Fidel2}) up to
the order $O(y^2)$:
\begin{equation}
{\cal F}(\rho(x-y), \rho(x+y)) = 1 - y^2\sum_{m,n} \frac{|\langle n|\rho'(x)|m \rangle|^2}
{[\rho_m(x)+\rho_n(x)]}, \qquad x,y \in R.
\label{Fidel3}
\end{equation}
Hence, by setting here $y=h/2$, from the definition (\ref{dchi2}) of the fidelity susceptibility, it follows that
\begin{equation}
\chi_{F}^{(2)}(\rho(x))= \frac{1}{2}\sum_{m,n} \frac{|\langle n|\rho'(x)|m \rangle|^2}
{[\rho_m(x)+\rho_n(x)]}.
\label{dchi3}
\end{equation}
At $x=0$ this expression reduces exactly to the spectral representation (\ref{FiSus2}) found in our work \cite{BT12} for
the zero-field one-sided fidelity susceptibility.

\section{Lower and upper bounds on the fidelity susceptibility}

Bounds on the fidelity susceptibility follow by applying elementary inequalities for $(x\coth x)^{-1}$ to the summand in expression (\ref{FiSus2}). In our paper \cite{BT12} an upper bound on $\chi_F(\rho)$ 
was obtained in the transparent form
\begin{equation}
\chi_F(\rho) \leq \frac{\beta^2}{4}(\delta S;\delta S)_0,
\label{FiSUpf}
\end{equation}
where $(\delta S;\delta S)_0$ is the Bogoliubov-Duhamel inner product of the self-adjoint operator
$\delta S$ with itself in the Gibbs ensemble with Hamiltonian $H(0)=T$.
Note that the right-hand side of the above inequality is proportional
to the initial  thermodynamic susceptibility:
\begin{equation}
(\delta S; \delta S)_0
= -\frac{N}{\beta}\frac{\partial^2 f[H(h)]}
{\partial^{2} h}\mid_{h=0}:=\frac{N}{\beta}\chi_{N},
\label{hi}
\end{equation}
where $f[H(h)]$ is the free energy density of the system described by the
Hamiltonian $H(h)$ and $\chi_{N}$ is the susceptibility with respect
to the field $h$.

On the other hand, by applying to the spectral representation for the fidelity susceptibility (\ref{FiSus2})
the elementary inequality $(x\coth x)^{-1} \geq 1 -(1/3)x^2$, we have obtained the following lower bound
\begin{equation}
\chi_F(\rho) \geq
\frac{\beta^2}{4}(\delta S;\delta S)_0 - \frac{\beta^3}{48}\langle [[S,T], S]\rangle_0.
\label{FiSusLo}
\end{equation}

The quality of the derived upper and lower bounds was tested in the simplest
case of a single spin in external magnetic field, subject to a transverse-field perturbation.
Finally, these bounds were applied to two many-body quantum-mechanical models: the
single impurity Kondo model and the Dicke model of superradiance.
In conclusion, our  lower (\ref{FiSusLo}) and upper (\ref{FiSUpf}) bounds indicate that for the detection of a second order phase transition, with diverging in the thermodynamic limit susceptibility,
the fidelity susceptibility per particle $\chi_F/N$ is as efficient as the usual
susceptibility  $\chi$. This conclusion is in conformity with the commonly accepted view that
quantum fluctuations are dominated by the thermal ones when $T_c >0$. However, one should keep in mind
that our results were derived under rather restrictive conditions on the spectrum of the Hamiltonian.

\section{Concluding remarks}

The infinite sets of generalized statistical mechanical inequalities presented in Section II provide upper bounds on the difference between the quadratic fluctuations of intensive observables expressed in terms of the corresponding Bogoliubov - Duhamel inner product and Gibbs average values of their commutator with the Hamiltonian. Such bounds are used, e.g., in the majorization technique
developed by Bogoliubov Jr. for the needs of the Approximating Hamiltonian method \cite{Bog72}.
A survey of inequalities used to solve problems arising in the Approximating Hamiltonian Method, along with their generalizations, is given in our paper \cite{BT11}. The results are illustrated by two types of exactly solvable model systems: one with bounded separable attraction and the other describing interaction of a boson field with matter.

In Section III,  some subtle points in the definition of the thermal fidelity susceptibility are discussed. The concept of fidelity susceptibility naturally appears as the fidelity's leading term in the perturbation expansion with respect to the infinitesimal deviation from a particular point of the parameter space. So it is possible to have different definitions depending on the way this point is approached. Two definitions are commonly considered: a one-sided second derivative with respect of the parameter distinguishing the two density matrices, or a symmetric two-sided derivative.
In the next Section IV, it is shown that the final result for the spectral presentation of the fidelity susceptibility does not depend on which of the
two definitions is used. While it seems intuitively reasonable, it is nevertheless important to prove this statement.

In Section  V, we have presented bounds on the fidelity susceptibility, a notion from the information theory, which are expressed in terms of
quantities from the statistical mechanics, thus emphasizing connection points between these two disciplines. An additional reason that stimulates this line of consideration of information-theoretic quantities is that the experimental setup for measuring
 thermodynamic quantities is well developed. Thus, estimation of  metric
 quantities with the aid of thermodynamic-based experiments seems to be very appealing.
 Note that the fidelity susceptibility reduces down to the Fisher information (for details see, e. g., \cite{ZVG07,ZPV08} and references therein) which
 provides another line of applicability of our results to quantum estimation theory.

We have shown that as far as divergent behavior  in the thermodynamic limit is considered, the fidelity
susceptibility $\chi_F $ and the  usual thermodynamic susceptibility $\chi$
are equivalent for a large class of models exhibiting critical behavior.
It remains for the future to study the effect of the degeneracy of the ground state, especially of a macroscopic one,
on the upper and lower bounds for the fidelity susceptibility.

{\bf Acknowledgement:}
The financial support of a collaboration grant of the Plenipotentiary Representative of the Government of Bulgaria at the Joint Institute
for Nuclear Research, Dubna, is gratefully acknowledged.

\end{document}